\documentclass[a4paper]{article}

\usepackage{INTERSPEECH2022}
\usepackage{algorithm}
\usepackage{algpseudocode}
\usepackage{amsmath}
\usepackage{hyperref}
\usepackage{multirow}
\usepackage{graphicx}
\usepackage{subcaption}

\title{SpeechEQ: Speech Emotion Recognition based on Multi-scale Unified Datasets and Multitask Learning}
\name{Zuheng Kang, Junqing Peng, Jianzong Wang$^*$\thanks{ $^*$Corresponding author: Jianzong Wang, jzwang@188.com}, Jing Xiao}
\address{
  Ping An Technology (Shenzhen) Co., Ltd.}
\email{\{kangzuheng896, pengjq, wangjianzong347, xiaojing661\}@pingan.com.cn}

\begin{document}

\maketitle
\begin{abstract}

  Speech emotion recognition (SER) has many challenges, but one of the main challenges is that each framework does not have a unified standard.
  In this paper, we propose SpeechEQ, a framework for unifying SER tasks based on a multi-scale unified metric.
  This metric can be trained by Multitask Learning (MTL), which includes two emotion recognition tasks of Emotion States Category (EIS) and Emotion Intensity Scale (EIS), and two auxiliary tasks of phoneme recognition and gender recognition.
  For this framework, we build a Mandarin SER dataset - SpeechEQ Dataset (SEQD).
  We conducted experiments on the public CASIA and ESD datasets in Mandarin, which exhibit that our method outperforms baseline methods by a relatively large margin, yielding 8.0\% and 6.5\% improvement in accuracy respectively.
  Additional experiments on IEMOCAP with four emotion categories (i.e., angry, happy, sad, and neutral) also show the proposed method achieves a state-of-the-art of both weighted accuracy (WA) of 78.16\% and unweighted accuracy (UA) of 77.47\%.

\end{abstract}


\section{Introduction}

Emotions, an ``implicit channel'' that transmits explicit messages \cite{cowie2001emotion}, play a vital role in human communication.
Speech is an important carrier of emotion and the easiest way to record information completely.
Modern people usually communicate by phone or through APPs.
Unfortunately, existing technologies lack standards for speech emotion information.
If designed a unified SER framework that automatically recognizes the human emotional state categories (ESC), such as happy, sad, anger, and its emotional intensity scales (EIS) expressed in natural speech, the intelligence systems can better perceive human thoughts and have better interactivity.

Recently, deep-learning-based approaches show great performance in extracting hidden speech information, including age \cite{si2022towards}, facial expression \cite{si2021speech2video} and emotion, etc.

The improvement of network architecture with recurrent neural networks (RNN) and attention mechanism \cite{li2021speech, peng2021efficient}, or analyzing and fusing the textual modal information \cite{chen2022key, singh2021multimodal} has demonstrated impressive results in solving SER problem.
However, since emotion is a complex psychological process, a single task may not be sufficient to capture complicated emotional features.

Using a transfer model for different tasks can also improve accuracy.
Because it essentially uses external knowledge to make the model more robust.
To capture emotional information, researchers combine the power of automatic speech recognition (ASR) with natural language processing (NLP) into the SER network structure to improve model performance \cite{padi2022multimodal, cai2021speech}.
However, their solutions are to fuse the information of different tasks in the last few layers (high-level features) of the network, which makes the information of different tasks difficult to communicate with each other.
In fact, the information of different tasks is intrinsically correlated with each other.
In SER, the speed and the way of speaking will affect the expression of human emotions.
Therefore, a unified model will comprehensively consider these factors to give a reasonable inference.

Multi-task learning (MTL) uses a shared backbone model to simultaneously optimize multiple objectives in different tasks.
The advantage comes from side information and cross-regularization of different tasks.
At the same time, joint optimization also brings challenges \cite{crawshaw2020multi}.
In SER, MTL provides an idea for the model to learn multimodal information simultaneously \cite{cai2021speech, li2019improved, nediyanchath2020multi}.
However, these methods use only one dataset for training and testing, which leads to a lack of model generalization.

Self-supervised learning utilizes multiple datasets to generate high-quality speech features \cite{eskimez2018unsupervised, si2021variational} for SER.
However, this method does not fully exploit hidden information in these datasets.

The released SER datasets have:
(1) content-dependent (such as IEMOCAP \cite{busso2008iemocap}) or content independent (such as RAVDESS \cite{livingstone2018ryerson}, ESD \cite{zhou2022emotional}, CREMA-D \cite{cao2014crema});
(2) contain EIS (such as IEMOCAP, CREMA-D, RAVDESS) or not (CASIA \cite{zhang2008design});
(3) lack a unified standard for the ESC in each dataset.

To find this unifying criterion, we took inspiration from psychologists' theories of emotion.
One of the most well-known theories is the emotional wheel proposed by Psychologist Robert Plutchik.
He theorized 8 emotions with 24 ``primary'', ``secondary'', and ``tertiary'' dyads \cite{plutchik1980general}.
Later, E. Cambria et al. \cite{cambria2012hourglass} updated his theory and created the hourglass of emotions - he divided emotion into 4 dimensions, but reversed the position of each emotion, such that the intensity of emotion on each dimension could be a numerical value.
These theories give us an idea to create a calculable metric for speech emotions.
And the main contributions of our works are as follows:

\begin{itemize}
  \item This paper proposed SpeechEQ, using a multi-scale unified metric, SpeechEQ Metric (SEQM), that unifies all frameworks. It can be trained with an MTL framework to simultaneously perform two emotion recognition tasks, ESC and EIS, and two auxiliary tasks of phoneme recognition and gender recognition;
  \item We build a Mandarin SER dataset - SpeechEQ dataset (SEQD) to demonstrate that using this metric, the emotion recognition accuracy of each Mandarin SER dataset can be improved;
  \item The effectiveness of this method is also verified on the public English SER dataset IEMOCAP;
\end{itemize}

\section{Methodology}

\subsection{Multiscale Unified Metric}

\begin{figure}[ht]
  \centering
  \includegraphics[width=0.45\textwidth]{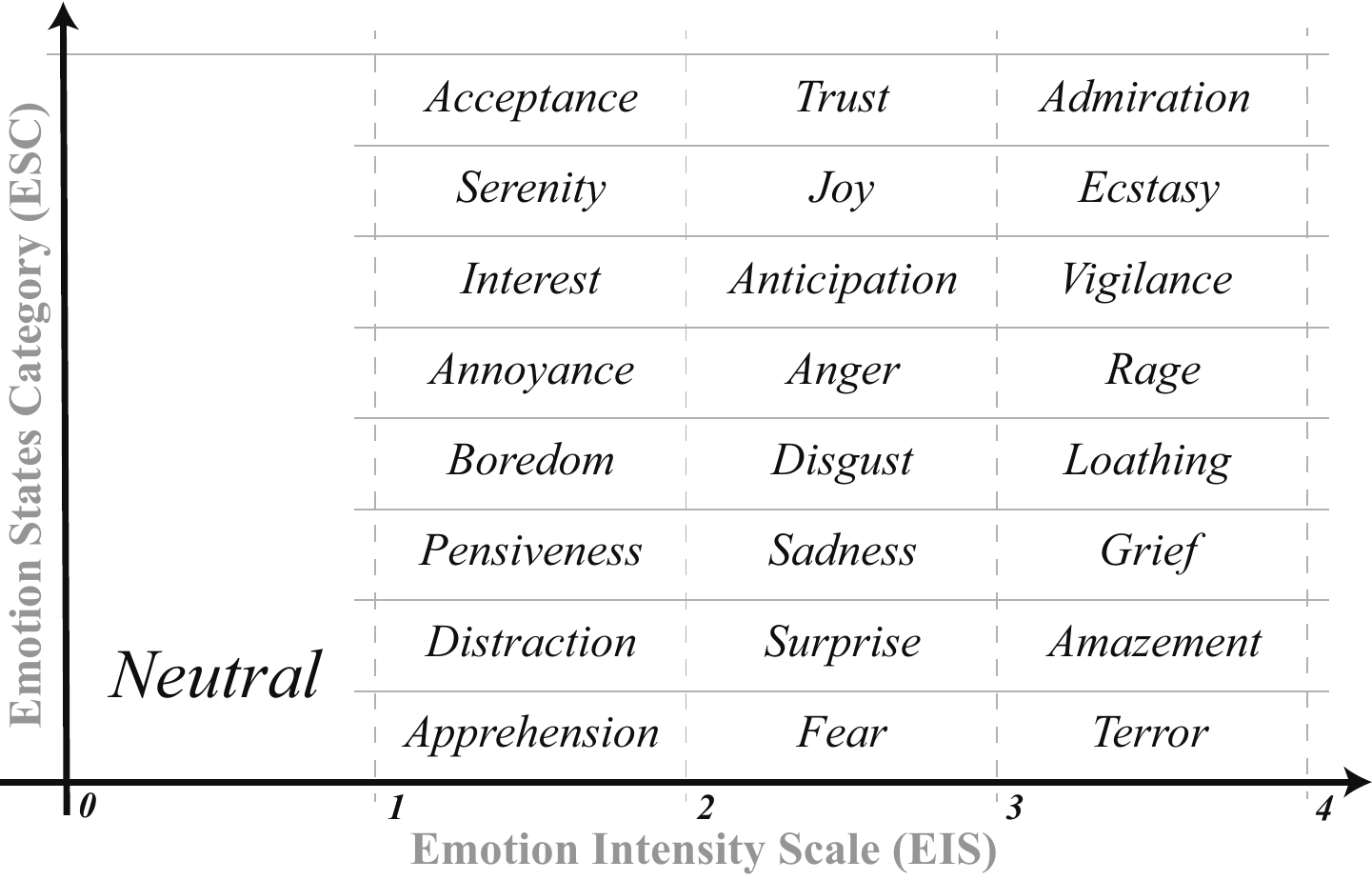}
  \caption{the SpeechEQ Metric for SER.}
  \label{fig_emotion}
\end{figure}

In order to unify different frameworks into a unified standard, we need a metric.
Inspired by the emotional hourglass theory, we divide human emotions into 8 ESC.
Then, use a numerical value to represent the intensity of the emotion: a value from 1 to 4 represents the EIS from low to high.
Finally, very weak emotions with low EIS values ranging from 0 to 1 are defined as the 9th category -- Neutral.
We name it, the SpeechEQ metric (SEQM), shown in Figure \ref{fig_emotion}.
With this metric, multiple datasets can be fused into a Multiscale Unified Dataset (MsUD) -- a label unification mechanism.

In modeling, since emotion categories are independent of each other, but the discrimination of emotional intensity is vague, sometimes indistinguishable, our proposed model will classify emotions into 9 categories and regress emotional intensity.
To create the gold label for training, if the EIS is divided into 3 levels (i.e., low, medium, high), the gold label will be set to 1.5, 2.5, and 3.5 (neutral is 0).
If with 2 levels, set them to 2 and 3 respectively.
If emotional intensity has been recorded in other forms, rescale to the range of 1 to 4.

\subsection{SpeechEQ Dataset}
To test the effectiveness of the SEQM, we constructed the SpeechEQ Dataset (SEQD), a content-independent Mandarin SER dataset, to help us improve the overall model performance in Mandarin SER tasks.
This dataset contains a total of 2.3 hours of speech, 1648 audio clips from 20 speakers (10 males and 10 females) at a sampling rate of 16kHz and a precision of 16 bits.
It was recorded with a Huawei phone in a medium-sized conference room at a signal-to-noise ratio (SNR) of about 20dB.
The speaker is about 20cm from the microphone, sitting in the center of the room.
See Table \ref{table_seqd} for details (Neutral is 73).

\begin{table}[ht]
  \centering
  \scriptsize
  \setlength{\tabcolsep}{2pt}
  \renewcommand{\arraystretch}{1}
  \caption{Number of utterances in SEQD with emotions and level (use its Medium emotion name to represent overall ESC).}
  \begin{tabular}{@{}lllllllll@{}}
    \toprule
                    & \textbf{Trust} & \textbf{Joy} & \textbf{Anticipation} & \textbf{Anger} & \textbf{Disgust} & \textbf{Sadness} & \textbf{Surprise} & \textbf{Fear} \\ \midrule
    \textbf{Low}    & 62             & 59           & 60                    & 72             & 66               & 61               & 61                & 72            \\
    \textbf{Medium} & 62             & 63           & 65                    & 77             & 73               & 77               & 63                & 63            \\
    \textbf{High}   & 70             & 59           & 60                    & 71             & 67               & 62               & 70                & 60            \\ \bottomrule
  \end{tabular}
  \label{table_seqd}
\end{table}

To build this dataset, each speaker independently wrote 3 to 5 sentences describing the emotion for each of the 25 emotions in the SEQM and performed it independently in the tone of talking to someone in everyday conversation.
Then, each utterance was judged independently by three judges.
If more than one judge judges the recorded speech emotion to be inaccurate, the speaker is notified to re-record until all three judges agree that the emotion is matched.

\subsection{Model Structure}

For the speech feature extraction, the structure of ECAPA-TDNN \cite{desplanques2020ecapa} has efficient design structures such as Res2Net \cite{gao2019res2net} and Squeeze Excitation blocks (SE) \cite{hu2018squeeze}.
However, it is originally for the speaker recognition task, which results in a smaller receptive field.
SER requires more frames to understand contextual information for more comprehensive reasoning, which needs a larger receptive field.

\begin{figure}[ht]
  \centering
  \includegraphics[width=0.43\textwidth]{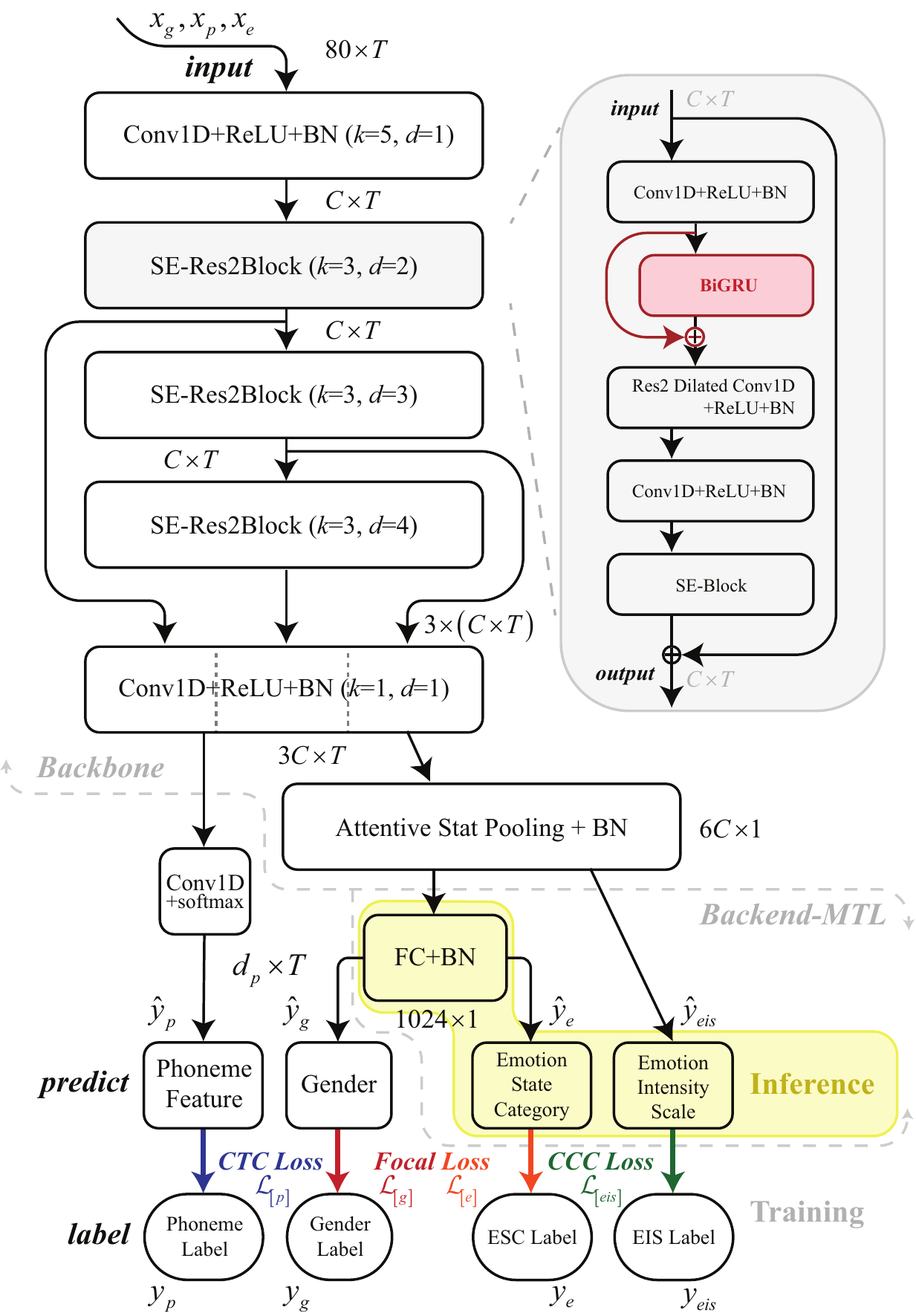}
  \caption{Network topology of the SpeechEQ framework.}
  \label{fig_net}
\end{figure}

In Figure \ref{fig_net}, in the backbone model, all structures are the same with ECAPA except Res-BiGRU (the residual structure of bidirectional GRU) in the red background, which has been added to SE-Res2Block that extends the receptive field.
The backend model is changed to the structure of MTL.
In our SER framework, in addition to predicting the category and intensity of emotions, the model also predicts the phonemes in speech and the gender of the speaker as auxiliary tasks.

\subsection{Training and Inference}

In the training phase, the generated predictions can have access to the labels via blue, red, orange, and green paths to generate loss functions respectively.
For each utterance in the training set, the blue path has access to the gold phoneme sequences labels $ y_p $ translated from text labels, the red path to the ground-truth gender labels $ y_g $, the orange path to the gold ESC labels $ y_e $ and the green path to the gold EIS labels $ y_{eis} $.

For phoneme sequence prediction (blue task), the Connectionist Temporal Classification (CTC) \cite{graves2006connectionist} as the metric between predicted phoneme probability vectors and the given gold phoneme sequence labels.
Assuming that the number of frames of speech features is $ T $, the phoneme feature dimension is $ d_p $.
Thus, the CTC loss for phoneme recognition task is:

\noindent
\begin{equation}
  \mathcal{L}_{\left[ p \right]}=\text{CTC}\left( y'_p,y_p \right) ,\,\,\text{where\,\,}y'_p=\text{softmax} \left( \hat{y}_p \right) \in \mathcal{P}^{T\times d_p}
  \label{eq_ctc}
\end{equation}

For classification of emotion and gender (red and organce task), the proposed classification tasks are trained with a focal loss (FL) \cite{lin2017focal} with a tunable parameter $ \gamma $.
Assume the gender and ESC feature dimension is $ d_g $ (=2) and $ d_e $ (=9).
The loss for ESC and gender recognition tasks is in Equation \ref{eq_fl}, the sign $ * $ denotes either $ g $ for gender, or $ e $ for ESC:

\noindent
\begin{equation}
  \mathcal{L}_{\left[ * \right]}=\text{FL}\left( \hat{y}_*,y_*,\gamma \right) ,\,\,\text{where\,\,}\hat{y}_*\in \mathcal{P}^{d_*}
  \label{eq_fl}
\end{equation}

For regression of EIS (green task), the Lin's Concordance Correlation Coefficient (CCC) \cite{lawrence1989concordance} is used.
Let $ \mathcal{L}_{\left[ eis \right]} $ denote the loss function for EIS:

\noindent
\begin{equation}
  \mathcal{L}_{\left[ eis \right]}=1-\text{CCC}\left( \hat{y}_{eis},y_{eis} \right) ,\,\,\text{where\,\,}\hat{y}_{eis}\in \mathcal{P}
  \label{eq_ccc}
\end{equation}

We introduce a hyper-parameter $ \alpha $, $ \beta $, $ \eta $ to combine four losses into a single one $ \mathcal{L} $.
$ \alpha $ controls the relative importance of the CCC loss for EIS, $ \beta $ of CTC loss for phoneme recognition, $ \eta $ of classification loss for gender recognition.
Finally, the model will be optimized with the following objective w.r.t $ \boldsymbol{\theta } $:

\noindent
\begin{equation}
  \underset{\boldsymbol{\theta }}{\min}\mathcal{L}=\mathcal{L}_{\left[ e \right]}+\alpha \mathcal{L}_{\left[ eis \right]}+\beta \mathcal{L}_{\left[ p \right]}+\eta \mathcal{L}_{\left[ g \right]}
  \label{eq_loss}
\end{equation}

In the training process, the ASR dataset, gender dataset, and emotion dataset are respectively converted to speech features and sent to the network, and their corresponding losses are calculated separately.
Finally, combine these losses into one loss to optimize the entire model.

However, in EIS tasks, some data do not have labels assigned.
For this type of data, here's a label ignore mechanism for unspecified EIS labels by setting a mask, i.e., temporarily setting them to a fake value of -1 when building the gold labels.
During training, these values can be captured and masked by replacing any label value in the mini-batch satisfied $ \hat{y}_{eis}<0 $ with the predicted such that there is no error between the predicted value and the gold label value.
To prevent the value from exceeding the value range, a clipping function is designed to clip the value to the range of 0 to 4, $ y_{eis}=\text{clip}\left( \hat{y}_{eis} \right) $.
The pseudo-code for the detailed training process is in Algorithm \ref{peudo_train}.

At inference time, two auxiliary tasks are abandoned, and only emotion recognition tasks with ESC and EIS are retained (yellow background in Figure \ref{fig_net}).
Next, the emotion category softmax can be replaced by the argmax operator, and select the most probable emotion class label as an output.

\subsection{Data Augmentation}

Since ASR data are recorded in a neutral tone at a moderate speech rate, yet the SER data contains various speech rates and intonations, we will use the package of librosa \cite{mcfee2015librosa} to adjust the pitch (randomly select pitch parameter from -3 to 3) and the speed (randomly select speed parameter from 0.8 to 1.3) of each utterance during training.
Additionally, all data used for training will be augmented by probabilistically adding noise with an SNR ranging from 10dB to 25dB, and reverberation.

\begin{algorithm}[t]
  \floatname{algorithm}{Algorithm}
  \renewcommand{\algorithmicrequire}{\textbf{Input:}}
  \renewcommand{\algorithmicensure}{\textbf{Output:}}
  \caption{SpeechEQ Training Policy}
  \label{peudo_train}
  \scriptsize
  \begin{algorithmic}[1]
    \Require Speech feature inputs $ x_g $, $ x_p $, $ x_e $ from gender, phoneme, and emotion datasets respectively
    \Ensure Corresponding predicted gender, phoneme, and ESC, and EIS labels $ y_g $, $ y_p $, $ y_e $, $ y_{eis} $
    \State initialize model parameters $ \text{M}\left( \boldsymbol{\theta ;}x \right) $
    \For{$ n\in \left[ 1,end \right] $}
    \For{$ i^{th} $ mini-batch in dataLoader }

    \State prepare input speech features $ x_{g}^{\left( i \right)} $, $ x_{p}^{\left( i \right)} $, $ x_{e}^{\left( i \right)} $
    \State prepare output gold labels $ y_{g}^{\left( i \right)} $, $ y_{p}^{\left( i \right)} $, $ y_{e}^{\left( i \right)} $, $ y_{eis}^{\left( i \right)} $
    \State get predictions from multi-datasets:
    \State \hspace{\algorithmicindent} $ \hat{y}_{g}^{\left( i \right)}\gets \text{M}\left( \boldsymbol{\theta ;}x_{g}^{\left( i \right)} \right) $,
    \State \hspace{\algorithmicindent} $ \hat{y}_{p}^{\left( i \right)}\gets \text{M}\left( \boldsymbol{\theta ;}x_{p}^{\left( i \right)} \right) $,
    \State \hspace{\algorithmicindent} $ \hat{y}_{e}^{\left( i \right)},\hat{y}_{eis}^{\left( i \right)}\gets \text{M}\left( \boldsymbol{\theta ;}x_{e}^{\left( i \right)} \right) $
    \State /* label ignore mechanism */
    \For{$ j^{th} $ element in mini-batch}
    \If{$ \hat{y}_{eis}^{\left( i,j \right)}<0 $ }
    \State $ y_{eis}^{\left( i,j \right)}\gets \text{clip}\left( \hat{y}_{eis}^{\left( i,j \right)} \right) $
    \EndIf
    \EndFor
    \State get FL loss for gender $ \mathcal{L}_{\left[ g \right]} $ using Equation \ref{eq_fl}
    \State get CTC loss for phoneme $ \mathcal{L}_{\left[ p \right]} $ using Equation \ref{eq_ctc}
    \State get FL loss for ESC $ \mathcal{L}_{\left[ e \right]} $ using Equation \ref{eq_fl}
    \State get CCC loss for EIS $ \mathcal{L}_{\left[ eis \right]} $ using Equation \ref{eq_ccc}
    \State merge losses with $ \mathcal{L} $ using Equation \ref{eq_loss}
    \State create and apply gradients with optimizer:
    \State \hspace{\algorithmicindent} update model parameters $ \boldsymbol{\theta ;} $
    \EndFor
    \EndFor

  \end{algorithmic}
\end{algorithm}

\section{Experiments}

In this paper, a total of 2 parts of the experiment will be performed.
The purpose of experiment 1 is to verify whether the MsUD and MTL are helpful to improve the accuracy of each dataset in Mandarin.
Experiment 2 confirms that our method is also helpful to the public English dataset -- IEMOCAP.

\subsection{Datasets and Metrics}
Multiple datasets will be used in these experiments, shown in Table \ref{table_dataset}.
The dataset for gender recognition uses the merging of all used SER datasets (training set).
For phoneme and emotion recognition tasks, they are determined by different experiments.

\begin{table}[t]
  \centering
  \scriptsize
  \setlength{\tabcolsep}{6pt}
  \renewcommand{\arraystretch}{1.1}
  \caption{Dataset details,
    abbreviation: N-neutral, A-angry, H-happy, S-sad, F-fear, D-disgust, Sr-surprise, Exc-excited.}
  \begin{tabular}{@{}llll@{}}
    \toprule
    \textbf{Dataset} & \textbf{Used Labels} & \textbf{Level}       & \textbf{Language} \\ \midrule
    SEQD             & SEQM emotions        & 3 levels             & Mandarin          \\
    CASIA            & A, F, H, N, S, Sr    & unspecified          & Mandarin          \\
    ESD              & A, H, N, S           & unspecified          & Mandarin+English  \\
    RAVDESS          & A, H, N, S           & 2 levels             & English           \\
    CREMA-D          & A, H, N, S           & 3 levels+unspecified & English           \\
    IEMOCAP          & A, H(+Exc), N, S     & continuous           & English           \\ \bottomrule
  \end{tabular}
  \label{table_dataset}
\end{table}

\textbf{Experiment 1:}
For the CASIA dataset, we perform a leave-one-fold-out ten-fold cross-validation experiment.
For the SEQD dataset, we give a fixed test set of 155 utterances.
The ESD dataset uses its given test set (Mandarin part, except Surprise).
The MsUD is built by the training part of SEQD, CASIA, ESD (Mandarin part, except Surprise).
The phoneme recognition training for Mandarin uses aishell-1 \cite{bu2017aishell}.
The phoneme gold label is built by the given transcript translated into
INITIALS and FINAL\_TONE3 by the package pypinyin, then converted to an integer through the phoneme dictionary (dictionary size is 202, with 21 initials, 36 finals, 5 tones, 1 silence).
The test metrics for the ESC are weighted accuracy (WA), unweighted accuracy (UA), and the EIS uses the mean square error (MSE).
The unweighted accuracy of important emotions (UAi) with neutral, happy, angry, and sad maybe also listed.
If the number of test samples for each class is the same (WA=UA), we denote it as Acc.
Finally, the average score of each cross-validation result is taken as the final performance score.
For datasets that do not have EIS labels (i.e., CASIA and ESD), it is trained using the label ignore mechanism.

\textbf{Experiment 2:}
For IEMOCAP, we select 4 emotions (i.e., angry, happy+excited, sad, and neutral) -- Table \ref{table_dataset}.
The experimental setup is the same as the cited system in Table \ref{table_result_iemo}.
We use the ``dominance'' value as its EIS, and rescale them from a range of 1 to 5, to a range of 1 to 4.
Then, randomly split the train/test into a 4:1 ratio and perform a 5-fold cross-validation.
The MsUD is built by important emotions of RAVDESS, CREMA-D, ESD (English part).
The phoneme recognition for English uses Librispeech \cite{panayotov2015librispeech}.
Transcripts can be directly translated into phonemes as gold labels via the dictionary CMUDict \cite{kominek2004cmu} (dictionary size is 85, with 84 phonemes and 1 silence).
Unspecified EIS labels are trained with the label ignore mechanism.
The test metrics are the same as in Experiment 1.

\subsection{Hyper-parameters}

For the speech feature, all datasets will be resampled to the sample rate at 16kHz and precision of 16 bits, and use the 80-dimensional Mel Filter-Banks feature extracted using the librosa package with frame size 25ms, hop size 10ms with Hamming window.
The channel parameter $ C $ in the convolutional layers for the proposed network is 256.
The dimension of the bottleneck in the SE-Block and the attention module is set to 128.
The scale dimensions in the Res2Block are set to 8.
The tunable parameter $ \gamma $ in FL is 10.

In the first stage, the parameters $ \alpha $, $ \beta $, $ \eta $ of the combined loss are initially set to be 1, 0.1, 0.1 respectively for fast convergence of each task.
All networks are optimized with the Adam optimizer \cite{kingma2014adam} with a learning rate of 1e-4 and weight-decay of 1e-5.
The mini-batch size for training is 32.

In the fine-tune stage, both of combined loss parameters $ \beta $ and $ \eta $ gradually decreased from 0.1 to 0.01. Furthermore, the learning rate decreased from 1e-4 to 1e-6.

\subsection{Evaluation Results}

\begin{table}[t]
  \centering
  \scriptsize
  \setlength{\tabcolsep}{3pt}
  \renewcommand{\arraystretch}{1.1}
  \caption{Ablation study of the proposed model on Mandarin datasets compared with baseline model.}
  \begin{tabular}{l|l|l|llll}
    \hline
    \multicolumn{1}{c|}{\multirow{2}{*}{\textbf{Experiment}}} & \textbf{CASIA}   & \textbf{ESD}     & \multicolumn{4}{c}{\textbf{SEQD}}                                                       \\ \cline{2-7}
    \multicolumn{1}{c|}{}                                     & \textbf{Acc}     & \textbf{Acc}     & \textbf{WA}                       & \textbf{UA}      & \textbf{UAi}     & \textbf{MSE}  \\ \hline
    Baseline                                                  & 93.27\%          & 89.43\%          & 55.83\%                           & 53.87\%          & 64.24\%          & 1.04          \\
    w/o MTL w/o MsUD                                          & 88.52\%          & 86.76\%          & 46.71\%                           & 45.04\%          & 52.64\%          & 0.98          \\ \hline
    \textbf{SpeechEQ}                                         & \textbf{96.45\%} & \textbf{93.25\%} & \textbf{66.14\%}                  & \textbf{65.25\%} & \textbf{86.91\%} & 0.68          \\
    w/o MTL                                                   & 95.32\%          & 92.33\%          & 62.47\%                           & 61.56\%          & 85.45\%          & \textbf{0.67} \\
    w/o MsUD                                                  & 93.15\%          & 90.46\%          & 52.24\%                           & 51.30\%          & 67.20\%          & 0.73          \\
    w/o MTL w/o MsUD                                          & 92.58\%          & 89.67\%          & 51.18\%                           & 50.16\%          & 55.01\%          & 0.71          \\ \hline
  \end{tabular}
  \label{table_result_seq}
\end{table}

\textbf{Experiment 1:}
The baseline model of CNN+RNN+Attention shares the same backbone model with \cite{li2019improved}.
Ablation studies are performed on the SpeechEQ framework with or without MTL or MsUD compared to the baseline model.

\begin{figure}[t]
  \centering
  \caption{EIS value distributions of Neutral, Low, Medium and High for training and testing set on SEQD (for SpeechEQ).}
  \begin{subfigure}{0.2\textwidth}
    \includegraphics[width=0.95\textwidth]{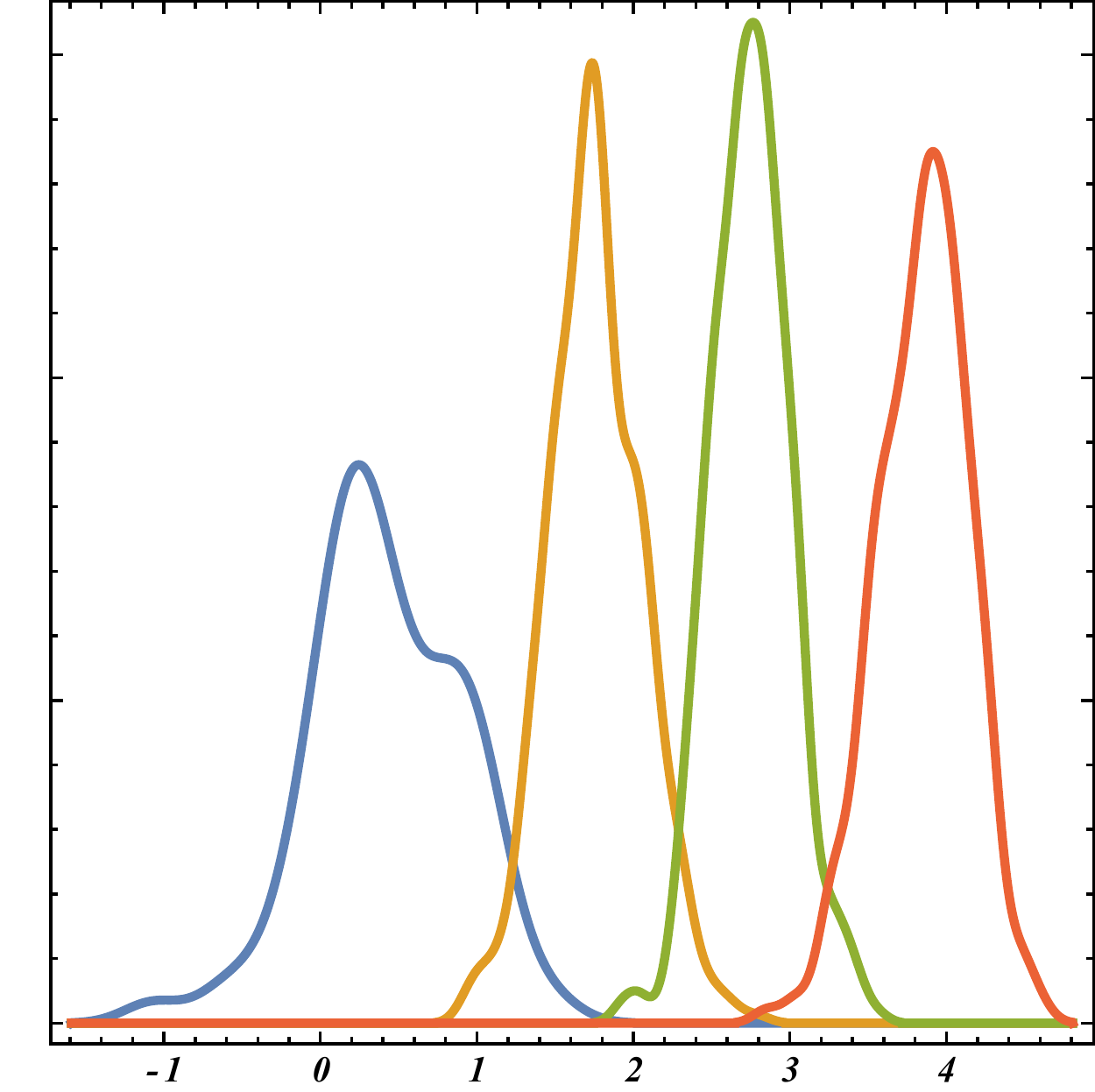}
    \caption{EIS for the training set.}
    \label{fig_eis_train}
  \end{subfigure}
  \begin{subfigure}{0.2\textwidth}
    \includegraphics[width=0.95\textwidth]{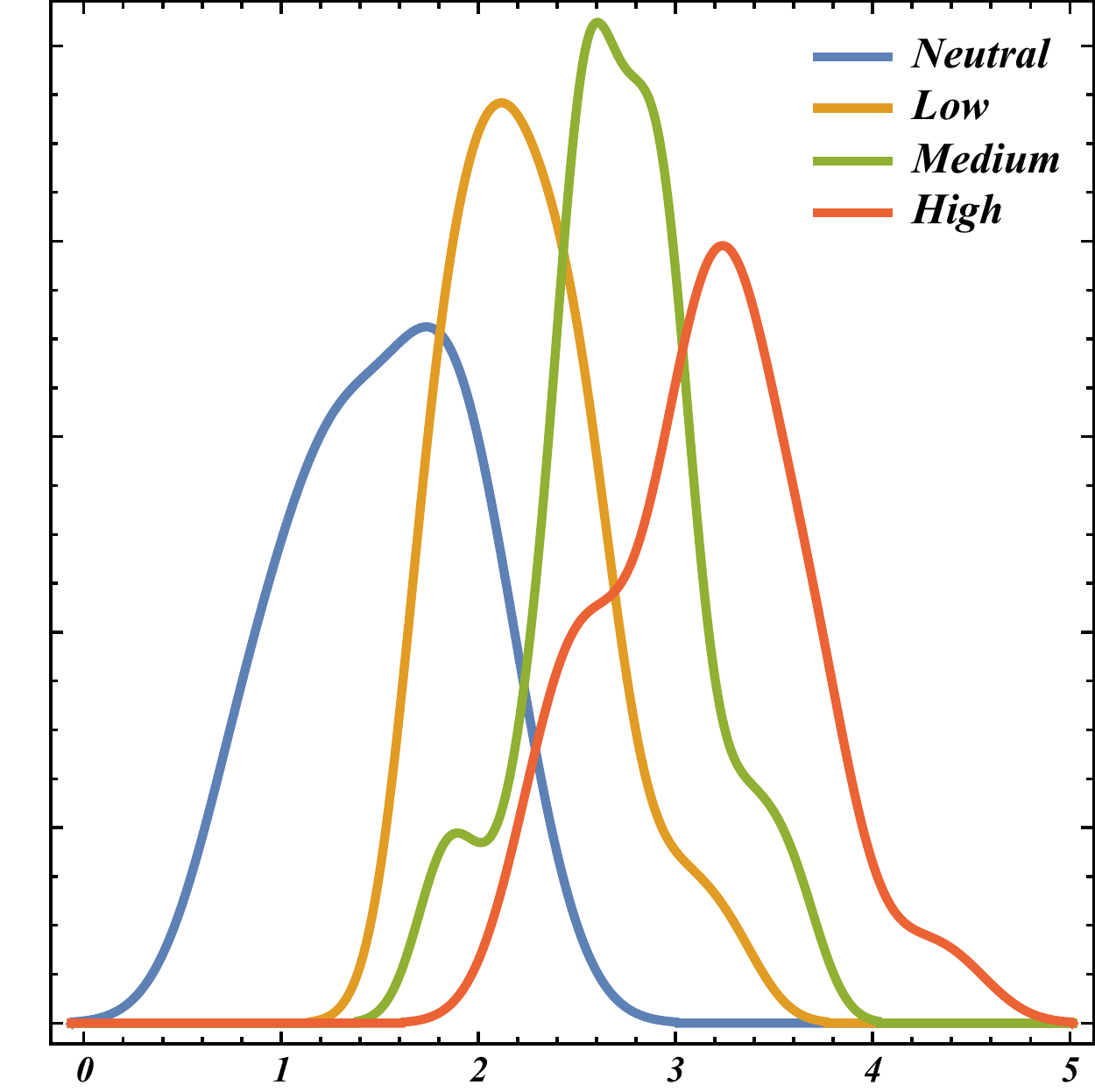}
    \caption{EIS for the testing set.}
    \label{fig_eis_test}
  \end{subfigure}
\end{figure}

The results are presented in Table \ref{table_result_seq}, and the EIS value distribution of Neutral, Low, Medium and High for training and testing set of SpeechEQ are shown in the Figure \ref{fig_eis_train}, \ref{fig_eis_test}.
Experimental results demonstrate that the feature extractor of SpeechEQ outperforms the baseline model with or without the help of MsUD and MTL.
Although adding auxiliary tasks of gender recognition and phoneme recognition improves the accuracy, training models on only a single dataset have limited improvement.
Without the use of MTL, the accuracy of the model is significantly improved after adding a MsUD.
Combining the two approaches, the accuracy of the model is further improved on both models.
From this result, it can be seen that training the model with either MsUD or MTL, or their combination, improves the accuracy of each dataset.
Meanwhile, the regression task of EIS will also be more convergent.

\textbf{Experiment 2:}
Compares the evaluation accuracy of training with or without MsUD and MTL, as well as other cited systems trained and tested only with IEMOCAP (results copied directly from the reference) in Table \ref{table_result_iemo}.
It is proved that the SpeechEQ framework is also applicable to the English dataset: using MsUD (with external data and labels) and MTL can significantly improve the ESC accuracy of the model, and further improve the state-of-the-art results by about 2\%.
At the same time, the EIS regression task is more convergent.

\begin{table}[t]
  \centering
  \scriptsize
  \setlength{\tabcolsep}{28pt}
  \renewcommand{\arraystretch}{1}
  \caption{Ablation study of the proposed model on IEMOCAP compared with cited SER results.
    The MSE for EIS at SpeechEQ and ``w/o MTL w/o MsUD'' are 0.74 and 0.81 respectively.}
  \begin{tabular}{@{}lll@{}}
    \toprule
    \textbf{Model}                             & \textbf{WA}      & \textbf{UA}       \\ \midrule
    CTC+Attention \cite{zhao2019attention}     & 67\%             & 69\%              \\
    BiGRU \cite{xu2020hgfm}                    & 66.6\%           & 70.5\%            \\
    MTL+Attention \cite{nediyanchath2020multi} & 76.4\%           & 70.1\%            \\
    Head Fusion \cite{xu2021head}              & 76.18\%          & 76.36\%           \\ \midrule
    \textbf{SpeechEQ (ours)}                   & \textbf{78.16\%} & \textbf{77.47\% } \\
    w/o MTL w/o MsUD                           & 73.20\%          & 74.62\%           \\ \bottomrule
  \end{tabular}
  \label{table_result_iemo}
\end{table}

\section{Conclusions}

In this paper, we propose SpeechEQ, a framework for unifying emotion recognition tasks:
(1) it unifies all SER frameworks with a multi-scale unified metric - SpeechEQ Metric;
(2) it can be trained on an MTL framework to perform not only ESC classification and EIS regression, but also auxiliary tasks of gender recognition and phoneme recognition.
Subsequently, we construct a Mandarin SER dataset - SEQD, to demonstrate the effectiveness of our method in improving model performance.
Finally, two experiments demonstrate the effectiveness of our method on both Mandarin and English SER tasks.

\section{Acknowledgement}

This paper is supported by the Key Research and Development Program of Guangdong Province under grant No.2021B0101400003.
Corresponding author is Jianzong Wang from Ping An Technology (Shenzhen) Co., Ltd (jzwang@188.com).

\clearpage
\bibliographystyle{IEEEtran}

\bibliography{ref}

\begin{thebibliography}{10}
\providecommand{\url}[1]{#1}
\csname url@samestyle\endcsname
\providecommand{\newblock}{\relax}
\providecommand{\bibinfo}[2]{#2}
\providecommand{\BIBentrySTDinterwordspacing}{\spaceskip=0pt\relax}
\providecommand{\BIBentryALTinterwordstretchfactor}{4}
\providecommand{\BIBentryALTinterwordspacing}{\spaceskip=\fontdimen2\font plus
\BIBentryALTinterwordstretchfactor\fontdimen3\font minus
  \fontdimen4\font\relax}
\providecommand{\BIBforeignlanguage}[2]{{%
\expandafter\ifx\csname l@#1\endcsname\relax
\typeout{** WARNING: IEEEtran.bst: No hyphenation pattern has been}%
\typeout{** loaded for the language `#1'. Using the pattern for}%
\typeout{** the default language instead.}%
\else
\language=\csname l@#1\endcsname
\fi
#2}}
\providecommand{\BIBdecl}{\relax}
\BIBdecl

\bibitem{cowie2001emotion}
R.~Cowie, E.~Douglas-Cowie, N.~Tsapatsoulis, G.~Votsis, S.~Kollias, W.~Fellenz,
  and J.~G. Taylor, ``Emotion recognition in human-computer interaction,''
  \emph{IEEE Signal processing magazine}, vol.~18, no.~1, pp. 32--80, 2001.

\bibitem{si2022towards}
S.~Si, J.~Wang, J.~Peng, and J.~Xiao, ``Towards speaker age estimation with
  label distribution learning,'' in \emph{ICASSP 2022-2022 IEEE International
  Conference on Acoustics, Speech and Signal Processing (ICASSP)}.\hskip 1em
  plus 0.5em minus 0.4em\relax IEEE, 2022, pp. 4618--4622.

\bibitem{si2021speech2video}
S.~Si, J.~Wang, X.~Qu, N.~Cheng, W.~Wei, X.~Zhu, and J.~Xiao, ``Speech2video:
  Cross-modal distillation for speech to video generation,'' in
  \emph{Proceedings of the Annual Conference of the International Speech
  Communication Association (INTERSPEECH)}, 2021.

\bibitem{li2021speech}
D.~Li, J.~Liu, Z.~Yang, L.~Sun, and Z.~Wang, ``Speech emotion recognition using
  recurrent neural networks with directional self-attention,'' \emph{Expert
  Systems with Applications}, vol. 173, p. 114683, 2021.

\bibitem{peng2021efficient}
Z.~Peng, Y.~Lu, S.~Pan, and Y.~Liu, ``Efficient speech emotion recognition
  using multi-scale cnn and attention,'' in \emph{IEEE International Conference
  on Acoustics, Speech and Signal Processing (ICASSP)}.\hskip 1em plus 0.5em
  minus 0.4em\relax IEEE, 2021, pp. 3020--3024.

\bibitem{chen2022key}
W.~Chen, X.~Xing, X.~Xu, J.~Yang, and J.~Pang, ``Key-sparse transformer for
  multimodal speech emotion recognition,'' in \emph{ICASSP 2022-2022 IEEE
  International Conference on Acoustics, Speech and Signal Processing
  (ICASSP)}.\hskip 1em plus 0.5em minus 0.4em\relax IEEE, 2022, pp. 6897--6901.

\bibitem{singh2021multimodal}
P.~Singh, R.~Srivastava, K.~Rana, and V.~Kumar, ``A multimodal hierarchical
  approach to speech emotion recognition from audio and text,''
  \emph{Knowledge-Based Systems}, vol. 229, p. 107316, 2021.

\bibitem{padi2022multimodal}
S.~Padi, S.~O. Sadjadi, D.~Manocha, and R.~D. Sriram, ``Multimodal emotion
  recognition using transfer learning from speaker recognition and bert-based
  models,'' \emph{arXiv preprint arXiv:2202.08974}, 2022.

\bibitem{cai2021speech}
X.~Cai, J.~Yuan, R.~Zheng, L.~Huang, and K.~Church, ``Speech emotion
  recognition with multi-task learning,'' in \emph{IEEE Conference of the
  International Speech Communication Association (INTERSPEECH)}, vol. 2021,
  2021.

\bibitem{crawshaw2020multi}
M.~Crawshaw, ``Multi-task learning with deep neural networks: A survey,''
  \emph{arXiv preprint arXiv:2009.09796}, 2020.

\bibitem{li2019improved}
Y.~Li, T.~Zhao, and T.~Kawahara, ``Improved end-to-end speech emotion
  recognition using self attention mechanism and multitask learning.'' in
  \emph{IEEE Conference of the International Speech Communication Association
  (INTERSPEECH)}, 2019, pp. 2803--2807.

\bibitem{nediyanchath2020multi}
A.~Nediyanchath, P.~Paramasivam, and P.~Yenigalla, ``Multi-head attention for
  speech emotion recognition with auxiliary learning of gender recognition,''
  in \emph{IEEE International Conference on Acoustics, Speech and Signal
  Processing (ICASSP)}.\hskip 1em plus 0.5em minus 0.4em\relax IEEE, 2020, pp.
  7179--7183.

\bibitem{eskimez2018unsupervised}
S.~E. Eskimez, Z.~Duan, and W.~Heinzelman, ``Unsupervised learning approach to
  feature analysis for automatic speech emotion recognition,'' in \emph{IEEE
  International Conference on Acoustics, Speech and Signal Processing
  (ICASSP)}.\hskip 1em plus 0.5em minus 0.4em\relax IEEE, 2018, pp. 5099--5103.

\bibitem{si2021variational}
S.~Si, J.~Wang, H.~Sun, J.~Wu, C.~Zhang, X.~Qu, N.~Cheng, L.~Chen, and J.~Xiao,
  ``Variational information bottleneck for effective low-resource audio
  classification,'' in \emph{Proceedings of the Annual Conference of the
  International Speech Communication Association (INTERSPEECH)}, 2021, p.~31.

\bibitem{busso2008iemocap}
C.~Busso, M.~Bulut, C.-C. Lee, A.~Kazemzadeh, E.~Mower, S.~Kim, J.~N. Chang,
  S.~Lee, and S.~S. Narayanan, ``Iemocap: Interactive emotional dyadic motion
  capture database,'' \emph{Language resources and evaluation}, vol.~42, no.~4,
  pp. 335--359, 2008.

\bibitem{livingstone2018ryerson}
S.~R. Livingstone and F.~A. Russo, ``The ryerson audio-visual database of
  emotional speech and song (ravdess): A dynamic, multimodal set of facial and
  vocal expressions in north american english,'' \emph{PloS one}, vol.~13,
  no.~5, p. e0196391, 2018.

\bibitem{zhou2022emotional}
K.~Zhou, B.~Sisman, R.~Liu, and H.~Li, ``Emotional voice conversion: Theory,
  databases and esd,'' \emph{Speech Communication}, vol. 137, pp. 1--18, 2022.

\bibitem{cao2014crema}
H.~Cao, D.~G. Cooper, M.~K. Keutmann, R.~C. Gur, A.~Nenkova, and R.~Verma,
  ``Crema-d: Crowd-sourced emotional multimodal actors dataset,'' \emph{IEEE
  transactions on affective computing}, vol.~5, no.~4, pp. 377--390, 2014.

\bibitem{zhang2008design}
J.~T. F. L.~M. Zhang and H.~Jia, ``Design of speech corpus for mandarin text to
  speech,'' in \emph{The Blizzard Challenge}, 2008.

\bibitem{plutchik1980general}
R.~Plutchik, ``A general psychoevolutionary theory of emotion,'' in
  \emph{Theories of emotion}.\hskip 1em plus 0.5em minus 0.4em\relax Elsevier,
  1980, pp. 3--33.

\bibitem{cambria2012hourglass}
E.~Cambria, A.~Livingstone, and A.~Hussain, ``The hourglass of emotions,'' in
  \emph{Cognitive behavioural systems}.\hskip 1em plus 0.5em minus 0.4em\relax
  Springer, 2012, pp. 144--157.

\bibitem{desplanques2020ecapa}
B.~Desplanques, J.~Thienpondt, and K.~Demuynck, ``Ecapa-tdnn: Emphasized
  channel attention, propagation and aggregation in tdnn based speaker
  verification,'' \emph{arXiv preprint arXiv:2005.07143}, 2020.

\bibitem{gao2019res2net}
S.-H. Gao, M.-M. Cheng, K.~Zhao, X.-Y. Zhang, M.-H. Yang, and P.~Torr,
  ``Res2net: A new multi-scale backbone architecture,'' \emph{IEEE transactions
  on pattern analysis and machine intelligence}, vol.~43, no.~2, pp. 652--662,
  2019.

\bibitem{hu2018squeeze}
J.~Hu, L.~Shen, and G.~Sun, ``Squeeze-and-excitation networks,'' in
  \emph{Proceedings of the IEEE conference on computer vision and pattern
  recognition}, 2018, pp. 7132--7141.

\bibitem{graves2006connectionist}
A.~Graves, S.~Fern{\'a}ndez, F.~Gomez, and J.~Schmidhuber, ``Connectionist
  temporal classification: labelling unsegmented sequence data with recurrent
  neural networks,'' in \emph{Proceedings of the 23rd international conference
  on Machine learning}, 2006, pp. 369--376.

\bibitem{lin2017focal}
T.-Y. Lin, P.~Goyal, R.~Girshick, K.~He, and P.~Doll{\'a}r, ``Focal loss for
  dense object detection,'' in \emph{Proceedings of the IEEE international
  conference on computer vision}, 2017, pp. 2980--2988.

\bibitem{lawrence1989concordance}
I.~Lawrence and K.~Lin, ``A concordance correlation coefficient to evaluate
  reproducibility,'' \emph{Biometrics}, pp. 255--268, 1989.

\bibitem{mcfee2015librosa}
B.~McFee, C.~Raffel, D.~Liang, D.~P. Ellis, M.~McVicar, E.~Battenberg, and
  O.~Nieto, ``librosa: Audio and music signal analysis in python,'' in
  \emph{Proceedings of the 14th python in science conference}, vol.~8.\hskip
  1em plus 0.5em minus 0.4em\relax Citeseer, 2015, pp. 18--25.

\bibitem{bu2017aishell}
H.~Bu, J.~Du, X.~Na, B.~Wu, and H.~Zheng, ``Aishell-1: An open-source mandarin
  speech corpus and a speech recognition baseline,'' in \emph{20th Conference
  of the Oriental Chapter of the International Coordinating Committee on Speech
  Databases and Speech I/O Systems and Assessment (O-COCOSDA)}.\hskip 1em plus
  0.5em minus 0.4em\relax IEEE, 2017, pp. 1--5.

\bibitem{panayotov2015librispeech}
V.~Panayotov, G.~Chen, D.~Povey, and S.~Khudanpur, ``Librispeech: an asr corpus
  based on public domain audio books,'' in \emph{IEEE international conference
  on acoustics, speech and signal processing (ICASSP)}.\hskip 1em plus 0.5em
  minus 0.4em\relax IEEE, 2015, pp. 5206--5210.

\bibitem{kominek2004cmu}
J.~Kominek and A.~W. Black, ``The cmu arctic speech databases,'' in \emph{Fifth
  ISCA workshop on speech synthesis}, 2004.

\bibitem{kingma2014adam}
D.~P. Kingma and J.~Ba, ``Adam: A method for stochastic optimization,''
  \emph{arXiv preprint arXiv:1412.6980}, 2014.

\bibitem{zhao2019attention}
Z.~Zhao, Z.~Bao, Z.~Zhang, N.~Cummins, H.~Wang, and B.~Schuller,
  ``Attention-enhanced connectionist temporal classification for discrete
  speech emotion recognition,'' \emph{Proceedings of the IEEE International
  Speech Communication Association (INTERSPEECH)}, 2019.

\bibitem{xu2020hgfm}
Y.~Xu, H.~Xu, and J.~Zou, ``Hgfm: A hierarchical grained and feature model for
  acoustic emotion recognition,'' in \emph{IEEE International Conference on
  Acoustics, Speech and Signal Processing (ICASSP)}.\hskip 1em plus 0.5em minus
  0.4em\relax IEEE, 2020, pp. 6499--6503.

\bibitem{xu2021head}
M.~Xu, F.~Zhang, and W.~Zhang, ``Head fusion: improving the accuracy and
  robustness of speech emotion recognition on the iemocap and ravdess
  dataset,'' \emph{IEEE Access}, vol.~9, pp. 74\,539--74\,549, 2021.

\end{thebibliography}

\end{document}